\documentclass[10pt, oneside]{article}   	% use "amsart" instead of "article" for AMSLaTeX format
\usepackage{geometry,color}                		% See geometry.pdf to learn the layout options. There are lots.
\usepackage{graphicx}				% Use pdf, png, jpg, or eps§ with pdflatex; use eps in DVI mode
\usepackage{amssymb,hyperref,color}
\usepackage[T1]{fontenc}
\usepackage[greek.ancient, english]{babel}
\usepackage{teubner}

\title{Majorana and the bridge between matter and anti-matter}
\author{Francesco Vissani\\
\small\sc INFN, Laboratori Nazionali del Gran Sasso}
\date{}							% Activate to display a given date or no date

\begin{document}
\maketitle

\begin{abstract}
This short essay aims to offer a discursive presentation of three scientific articles by Ettore Majorana 
highlighting the fundamental importance of one of them - the last one - 
for the investigation of the intimate constitution of matter.
The search for evidence to support Majorana's thesis is the prime motivation of the conference  
``Multi-Aspect Young Oriented Advanced Neutrino Academy'' at the G.P.~Grimaldi Foundation in Modica, Sicily\footnote{Essay 
based on the minutes of the speech on 4 July 2023 addressed to the citizens as an introduction to the conference. 
Il Nuovo Cimento {\bf C6} article 351 (2024).} %;  doi: \url{10.1393/ncc/i2024-24351-1}.}.
\end{abstract}

%{\Huge  REFERENZE: 
%introduci 3 ref. di Majorana;
%Dirac Nobel;
%Gamow;
%Fermi 1932;
%anderson nobel;
%chadwick nobel;
%fermi 1933;
%Furry 1938;
%Thomson nobel;
%Dirac sea;
%Pauli 1941;
%Stuckelberg;
%Feynman la ragion delle anti-particell;
%mia cagata;
%pauli 1936;
%pauli weisskopf 1934
%Einstein e aether
%}

\parskip0.5ex
\section{Introduction}
It is difficult to talk about Majorana in  his native land, Sicily; and this is so 
not only because Majorana's science is hard to explain (as it actually is), 
certainly not because Antonino Zichichi founded a scientific centre dedicated to him in Erice, 
and even less so because every Sicilian of good culture has an idea about his story;
but also and simply because we are in a special place for the very origin of science.
We are in the middle of Magna Graecia. 

%\textgreek{abcdefghijklmnopqrstuwyxz}

Modica (\textgreek{M'otouka}) is more or less equidistant from Catania  (\textgreek{Kat'anh}), Majorana's birthplace, and from Agrigento  (\textgreek{'Akr'agas}), the birthplace of Empedocles, the philosopher who began the line of thought on the constitution of matter that led to chemistry.
In Lentini (\textgreek{Leont\~inoi}), 50 km from here, was born Gorgias, possibly the finest of Greek sophists,
who had Empedocles himself as a mentor\footnote{This considerations suggest that the exercise of rational thought 
in the public arena and the practice of science, far from being separate from each other, were originally very close.
It is curious to observe some parallelism with Abdera (\textgreek{'Abdhra}), where 
 Democritus (who spoke of atoms but also of mathematics, ethics and much more)
and Protagoras were both born.}.
In the immediate surroundings of Modica is also Siracusa  (\textgreek{Sur'akousai}), %Συράκουσαι 
the birthplace of Archimedes, the greatest mathematician of the period 
in which Greek science reached its zenith.

\bigskip
To speak of Majorana is to speak of such a person: an outstanding mathematician and a very deep thinker,
of great ethical and scientific rigour. His legacy consists of a dozen scientific articles, some on molecules and atoms, 
others on their constituents: nuclei and subatomic particles \cite{mu}. 
We will discuss this second group of works: there are three in all \cite{m1,m2,m3}. 
%
%\bigskip
%Parlare di Majorana  \`e parlare di una persona di questo livello: un matematico eccezionale ed un pensatore assai profondo,
%di grande rigore etico e scientifico. La sua eredit\`a consiste in una decina di  articoli scientifici \cite{mu}, alcuni su molecole ed atomi, 
%altri sui loro costituenti: nuclei e particelle subatomiche. 
%Parleremo proprio di questo secondo gruppo di lavori: son tre in tutto \cite{m1,m2,m3}. 
%

\begin{figure}[t!]
\centerline{\includegraphics[width=0.8\textwidth]{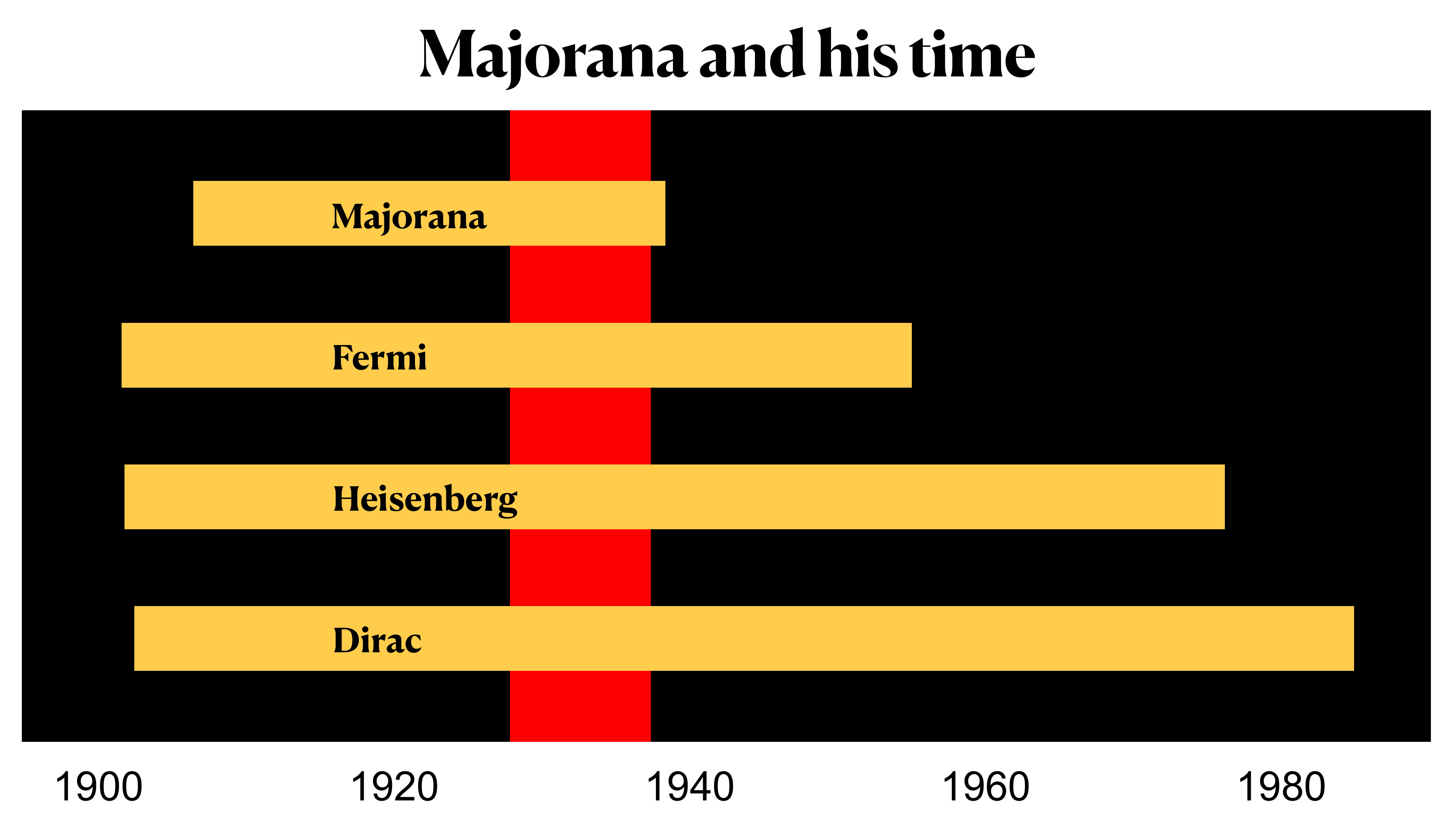}}
\caption{\small\it  Comparison of the essential biographical data of Ettore Majorana and three scientists, 
who have been particularly important to him. The red band indicates the period in which Majorana 
published the three works described in the text.\label{fig1}}
\end{figure}

We will try to explain 
why the last of these writings is still very relevant today.
It describes a new conception of particles, which leads one to think of anti-matter
in a more satisfactory and better way than the previous one: It is the way that all scholars of our time know and use. 
We will discover that Majorana's approach has led us to a different idea 
of certain particles of matter without electric charge, the {\em neutrinos;} 
an idea that many find convincing,  which comfortably fits into established theoretical views 
 and for which several laboratories around the world have set to work, to find observational evidence.

And it is not just a hypothesis that can be verified in the laboratory, but a 
a vision with a deep philosophical meaning. 
If Majorana is right, these very special neutral particles - which we will discuss later - could play no less than 
the role of a bridge between matter and anti-matter\footnote{Allow me to note that my interest 
in Ettore Majorana was born more than 30 years ago, from this very work of his, and is far from exhausted.},
 causing (though very rarely) processes in which electrons are created.

\medskip

Majorana confronted many of his contemporaries, in particular  two famous scientists slightly older than him, Fermi and Heisenberg - 
the only people who could vaguely give the impression of having played the role of advisor for him\footnote{The impression is that Majorana considered Fermi and Heisenberg as colleagues
with which it {\em makes sense} to deal.  It is less easy to guess what his perceptions were from a psychological point of view; 
 but  
 the figures of these two authoritative scientists, and the environments in which they met, have played 
 very different roles for Majorana.  Overall, I feel their relationship should be understood
like that among Greek philosophers: valuable pupils explore radical modifications of their mentor's theory.}.
Another contemporary with whom he establishes 
a kind of long-distance duel, no less intense and remarkable for that, 
is the brilliant British physicist Paul Dirac. 
Figure~\ref{fig1} offers a visual summary 
of their main biographical data.
Other important figures for Majorana that we will meet were Pauli, Weyl, Wigner,  
Gentile Jr - his colleague and friend  % \cite{gr} 
- and also 
the physicists of the Roman entourage, including of course the other 
 guys from Via Panisperna - the headquarters of the working group coordinated by Fermi.

\bigskip

I begin by recalling the theories of Dirac 
that invite Majorana to reflections, that will mature into fundamental scientific advances (Sect.~\ref{e-s2}). 
I then briefly describe the content of the three aforementioned works  (Sect.~\ref{e-s3}), and conclude
by illustrating the impressive and enduring value of the last of these  (Sects.~\ref{e-s4} and~\ref{e-s5}). 
I will only quote a few essential references 
relegating to the last pages some appendices  (Sect.~\ref{e-sa}), in which I have included 
supplementary material of various kinds that might prove useful to interested readers.

Before we get into the heart of the matter, I would like to present a few proposals for further study; very few indeed,
considering the nature of these notes.
I recommend reading the collection of letters, 
documents and testimonies
by Erasmo Recami~\cite{eras}, from which I have taken the passages reported 
 in the appendix~\ref{app:gent}.
On the science of Ettore Majorana - which is the theme I focus on - I commend the beautiful 
essay by Salvatore Esposito~\cite{salv} deep and complete, even if introductory in nature;
I also suggest reading the proceedings of the conference held in Catania 
a hundred years after his birth~\cite{conv}.
I will only touch marginally on Majorana's personal vicissitudes, 
but I cannot fail to mention at least Amaldi's recollection~\cite{ama}, 
the literary reconstruction by Leonardo Sciascia~\cite{leo}  (quite famous in Italy)
and two works by Guerra and Robotti~\cite{gr}; it might also be worth taking a look 
 of a recent essay collecting and examining numerous hypotheses put forward on his disappearance~\cite{gent}.

%Prima di entrare nel vivo, presento qualche proposta di approfondimento;  davvero poche,
%in considerazione del carattere di queste note.
%Raccomando la lettura della raccolta di lettere, 
%documenti e testimonianze
%di Erasmo Recami~\cite{eras}, da cui ho tratto i passaggi riportati 
% in appendice~\ref{app:gent}.
%Sulla scienza di Ettore Majorana  - che 
%\`e il tema su cui mi concentro -  caldeggio il bellissimo 
%saggio  di Salvatore Esposito~\cite{salv} profondo e completo anche se a carattere introduttivo;
%suggerisco anche la lettura degli atti della conferenza tenuta   a Catania 
%a cento anni della sua nascita~\cite{conv}.
%Toccher\`o solo marginalmente le vicende personali di Majorana, 
%ma non posso non segnalare almeno il ricordo di Amaldi \cite{ama},
%il celebre libro di Leonardo Sciascia~\cite{leo} 
%e due opere di Guerra e Robotti~\cite{gr}; potrebbe anche valer la pena di prendere visione 
% di un recente saggio che colleziona ed esamina numerose ipotesi avanzate sulla sua scomparsa~\cite{gent}.

%{\color{red} [[bibliografia essenziale: lavori raccolti,
%Salvatore sulla scienza, Recami sull'epistolario. Segnaliamo poi 
%un curioso lavoro raccoglie le numerosissime ipotesi sulla scomparsa di Majorana, di cui non parlaremo; vorremmo limitarci, per quanto possibile,  
%alla scienza. Convegno per il centenario. Sciascia che segue Zichichi 1972
%
%https://people.na.infn.it/~sesposit/MajoranaSite/documents/PresentazVolZanichelli.pdf

%
%confessare il perche' conosco Majorana 
%]]}
%

\section{Dirac: a doubt, the strange way out, success}\label{e-s2}
Dirac's most famous work is entitled
{\em The quantum theory of the electron} and is from 1928. The main result of physics
discussed there is the following: starting from the new wave-equation he obtained, 
Dirac provides excellent arguments for
the existence of the spin of the electron, showing how it is associated with a sort of small magnet 
which interacts with the magnetic field exactly as indicated by the experiments.
Any high school student, who has any familiarity with the theoretical concepts of chemistry, 
 knows that spin is a very important thing:  
 Dirac was onto something big.
 
%Il pi\`u famoso lavoro di Dirac si intitola 
%{\em La teoria quantistica dell'elettrone} ed \`e del 1928. Il principale  risultato di fisica 
%l\`i discusso \`e il seguente: 
%partendo dalla  nuova equazione d'onda ottenuta da lui stesso trovata, Dirac fornisce ottimi argomenti per 
%l'esistenza dello spin dell'elettrone,  mostrando come esso sia associato ad  una  sorta di piccola calamita, 
%che interagisce col campo magnetico  esattamente come indicato dagli esperimenti.
%Ora, qualsiasi studente  di scuola superiore che abbia  qualche familiarit\`a con i concetti della chimica 
%sa che lo spin \`e una cosa importantissima; quindi Dirac aveva qualcosa in mano. 

Perhaps the same student will also have heard of an even more impressive discovery, that of antimatter.
The starting point is again the same work, but those who read it will struggle to find even a word about it. 
It is a result that will require further reflection, and it will be another three years before Dirac claims that his own equation predicts the existence of \textbf{anti-electrons}. This is how he introduces it \cite{de31}:
\begin{quote}
{\sf \small ``a new kind of particle, unknown to experimental physics, having the same mass and opposite charge to an electron. We may call such a particle an anti-electron''. 

}
\end{quote}
The following year (1932), when particles corresponding to these characteristics were actually observed, it was a triumph for Dirac, who entitled his Nobel Prize lecture of December 1933 {\em Theory of electron and positron} \cite{nobeldirac}.
  In fact, the scientist who first observed the new particles - Anderson - had called them \textbf{positrons} \cite{andy}, since they have a positive charge.
  
%Quando nel 1932 (l'anno dopo) si osserveranno davvero delle particelle che corrispondono a queste caratteristiche,
%sar\`a un trionfo per Dirac, che intitoler\`a 
% la sua lezione per il premio Nobel del dicembre 1933 {\em Teoria dell'elettrone e del positrone}
% \cite{nobeldirac}. 
% Infatti, lo 
% scienziato che per primo aveva osservato le nuove particelle - Anderson - le aveva chiamate \textbf{positroni} \cite{andy}, siccome hanno carica positiva. 
 
 Today we use the names \textbf{anti-electron} and 
 \textbf{positron} interchangeably, as we have in the meantime become convinced that theory and observation are in neat correspondence.
  But in order to really understand the story, it is better to delve into these two concepts and their connection; we will do this later.
See the appendix~\ref{app:n} for a systematic discussion of the names relevant to the discussion.
 
% Oggi usiamo  indifferentemente i nomi
%anti-elettrone e positrone, in quanto ci siamo nel frattempo convinti che la teoria e l'osservazione siano in stretta corrispondenza;
%ma per capire davvero la storia, conviene approfondire meglio il nesso e lo facciamo qua sotto.
%Si veda l'appendice~\ref{app:n} per una disamina  sistematica dei  nomi rilevanti per la discussione.

\subsection{Dirac's reasoning}\label{fishy}
Dirac's success is indisputable and is still celebrated today.
But what are the arguments that he used to formulate his prediction?
I present them here, by detailing the conceptual steps he describes in~\cite{nobeldirac}:
\begin{itemize}
\item In the Dirac wave-equation there are solutions that would seem to describe {\em electrons with negative energy}; and it is not just  matter of a few states, but an infinite number of states with ever lower energies, never ending.
\item Let us now think of an electron revolving around the nucleus of an atom. Electrons, being electrically charged, can behave like antennae that emit radiation. The existence of strange states with `negative energies' would cause a disastrous race of electrons towards lower and lower energies: each atomic electron would have to continuously emit radiation, going lower and lower in energy. Since there appears to be no minimum energy value, each atom would continuously radiate energy and their very existence would appear to be threatened.
\item Something urgently needed to be rethought! And precisely in this regard, Dirac put forward a proposal of his own, which aimed to make usable the wave-equation that he had devised and which (unlike the proposed interpretation we are about to describe) we still use.
\item Previous observations and theories had established that the electrons around atoms occupy the permitted states {\em at most only once:} I am talking about the Pauli exclusion principle discovered in 1924 (see appendix~\ref{app:n}), 
which again I think is something we all heard about in high school\footnote{We should not forget the motivations and difficulties that Pauli had to overcome to be able to formulate it. Unfortunately, scholastic lectures do not usually help to appreciate this type of considerations, a circumstance which exposes students to the risk of believing that ideas arise from phantom geniuses, or let us say it, by magic (even though Pauli had what it took to play the role of the magician). For readers interested in learning more about it, I would like to point out~\cite{ppp}.}.
This principle suggests a very strange way out  to Dirac: he imagines that all states with negative energy
are already occupied. \newline
This is the {\bf\em Dirac sea hypothesis.}
\end{itemize}
In short, according to Dirac, we would all be immersed in an ocean of electrons with negative energies, completely uniform, and of which we could not be aware, like a fish swimming in the sea, far from the bottom and the surface: a rather disturbing idea. The great Soviet scientist George Gamow explained this strange possibility in the clearest possible way in his splendid book {\em Thirty Years That Shook Physics}~\cite{30yr}  (which I recommend to those who do not already know it). The chapter devoted to Dirac begins with a drawing by Gamow: a portrait of the great English scientist, intensely engaged in reflection in the depths of the sea and with a cute fish in front of his face.

%In short, according to Dirac we would all be immersed in an ocean
%of electrons with negative energy, completely uniform, and of which  we could not be aware, 
%as  a fish swimming in the sea,  distant  from the bottom and from the surface: a rather disturbing idea. 
%The great Soviet scientist and populariser George Gamow explained this odd possibility as clearly as possible in his wonderful book {\em Thirty Years That Shook Physics}~\cite{30yr} (which I recommend to those who do not already know it). The chapter dedicated to Dirac begins with a hand-made portrait of the great English scientist, intensely busy reflecting in the depths of the sea and with a cute fish right in front of his face.

%Insomma secondo Dirac saremmo tutti immersi in un oceano 
%di elettroni con energia negativa, del tutto uniforme, e del quale  
%non avremmo modo di renderci conto, come  chi si trovasse a  nuotare restando distante sia  
% dal fondo  che  dalla superficie: un'idea piuttosto inquietante.  Il grande scienziato e divulgatore sovietico George Gamow 
%l'ha spiegata nel modo pi\`u chiaro possibile nel suo meraviglioso libro {\em Trenta anni che sconvolsero la fisica}~\cite{30yr} (che raccomando a chi non lo conosca gi\`a). Il capitolo dedicato a Dirac inizia proprio con un ritratto dello scienziato inglese, intensamente impegnato a riflettere nel profondo il mare e con un bel pesce proprio davanti alla faccia. 

Fishy jokes aside,
up to the point of the discussion, we have limited ourselves to seeing how the disaster is avoided, or, if we prefer to be precise, how we try to avoid a radical inconsistency between the conceptual scheme adopted and some known facts, such as how the existence of atoms, 
which in Dirac's time was already indisputable\footnote{However, it is amusing to note that the Nobel Prize in Physics, awarding Jean Perrin definitive proof of the existence of molecules, dates back to 1926 - less than a century ago.}.
%\footnote{Si tratta di fatti molto semplici, del tipo: gli atomi esistono e sono stabili, se non perturbati.}.

\paragraph*{A prediction from Dirac's interpretation}
Let us now see how Dirac conceived anti-matter in his thought pattern. Every now and then, a quantum of light with particularly high energy could extract an electron from the Dirac sea, thus producing a ``normal'' one (=with positive energy). This extraction would be accompanied by the formation of a hole in the sea.
The lack of an electron of negative energy, with a given charge and momentum, would manifest itself as an increase in the energy of the sea, but also in the electric charge and even in the momentum.

In short, everything would lead us to believe that a particle has been created with a charge opposite to that of the electron, that moves in the opposite direction to the hole that has formed. This is the content of the so-called ``hole theory'', which as we have seen relies heavily on the idea that the Dirac sea exists. {\em Dirac's anti-electron, therefore, is nothing other than the lack of a negative energy electron in the sea:} a hole, in fact. 
The content of the 1931 document cited above \cite{de31}
consists precisely of this.

%Vediamo ora come Dirac concep\`i l'anti-materia nel suo schema di pensiero. Ogni tanto, un quanto di luce dotato di energia particolarmente alta potrebbe estrarre un elettrone del mare di Dirac, producendone cos\`i uno ``normale'' (di energia positiva). Questa estrazione sarebbe accompagnata dalla formazione di un buco nel mare.
%La mancanza di un elettrone di energia negativa, con data carica ed quantit\`a di moto, si manifesterebbe come un aumento dell'energia del mare, ma anche della carica elettrica, e persino della quantit\`a di moto. 

%In breve, tutto porterebbe a  credere che si \`e creata una particella con carica opposta a quella dell'elettrone, e che si muova in direzione opposta alla lacuna che si \`e formata. Questo \`e il contenuto della cosiddetta ``teoria del buco'', che come abbiamo visto si appoggia pesantemente all'idea che esista il mare di Dirac.  {\em L'anti-elettrone di Dirac, dunque, non \`e altro che  la mancanza di un elettrone nel mare:} un buco, per l'appunto.

\subsection{Back story and comments}
Since I have insisted a bit on the considerations of natural philosophy, let me observe that in Dirac's approach the electrons are comparable to the atoms of the ancient Greek thinkers: that is, eternal and immutable particles. We can extract them from the sea, or put them back in their place; but creating them, no: that  would not be possible. In other words, Dirac did not think as we think today, in terms of the creation of an electron-positron pair: for him, electrons were the same type of bulwark of the existence that Leucippus and Democritus had conceived, and on which they and many others after them would ponder\footnote{For anyone who doubts the value of this approach to the discussion, allow me to quote the words with which Maxwell (Dirac's compatriot) concludes the introduction to his course in 1871:
\begin{quote}
{\sf ``It has been asserted that metaphysical speculation is a thing of the past, and that physical science has extirpated it. The discussion, of the categories of existence, however, does not appear to be in danger of coming to an end in our time, and the exercise of speculation continues as fascinating to every fresh mind as it was in the days of Thales.''} \end{quote}}.

From this point of view, it is also interesting to recall a previous version of Dirac's theory, which was promptly set aside after the criticisms received from Oppenheimer and Tamm in 1930. In this ambitious speculation, Dirac ventured to identify the ``holes'' with the protons themselves. If it could ever work, this scheme would have encompassed {\em all}  matter  particles known at the time - electrons and protons - within a single system of thought; but unfortunately, if taken seriously, it would have led to the almost instantaneous disintegration of matter itself.

At this point in history, another humorous episode occurred: Wolfgang Pauli, known not only for being a profound scientist, but also for having a very sharp tongue, decided to expound another principle for the occasion (which was called ``Pauli's second principle''): every new speculative theory had first to be applied to the very people who had proposed it, and in this way there would be no need to listen to Dirac.

%
%Da questo punto di vista, \`e  anche interessante ricordare una versione precedente della teoria di Dirac, che  
%venne prontamente accantonata dopo le critiche ricevute da  Oppenheimer e Tamm nel 1930. In questa  ambiziosa speculazione, Dirac  azzard\`o l'identificazione dei ``buchi''  con i protoni stessi.  Se mai avesse potuto funzionare, questo schema avrebbe spiegato {\em tutte} le particelle di materia  note  all'epoca - elettroni e protoni - all'interno di un unico sistema di pensiero; ma purtroppo, se preso seriamente, esso avrebbe comportato la disintegrazione quasi istantanea della materia stessa. 
%
%Ci furono a questo punto dei risvolti umoristici: Wolfgang Pauli, noto  non solo per essere  uno scienziato profondo ma anche per avere una lingua assai tagliente, avanz\`o nell'occasione un altro principio (che venne chiamato ``secondo principio di Pauli''): ogni nuova teoria speculativa dovrebbe essere  preliminarmente applicata proprio a colui che la propone...   

\section{Majorana's last three works}\label{e-s3}

\subsection{The first confrontation with Dirac}\label{f1}
Dirac's ideas provoke widespread debate.  Hermann Weyl, a great mathematician, is  
among the first to intervene already in 1929,  proposing a different and simpler wave-equation \cite{weyl}, which however
seems unusable\footnote{The Weyl equation was put aside for the moment, since no way was found to describe the mass of the electron by means of it.}.
Instead, the reactions of the great physicists of the time are mixed; all those linked to Germany (including Bohr, Heisenberg, Pauli, Landau and many others) but also Rutherford and few more feel uncomfortable with the idea of Dirac's sea, and they make no secret of it.

It is interesting that Enrico Fermi distances himself from these positions. 
For example, he wrote a famous work on {\em Quantum theory of radiation}
\cite{fermirmp} early 1932, where he illustrated his effective way of putting Dirac's theory to work\footnote{To be precise, 
he also criticizes the version in which the holes are thought of as protons, but without saying anything against the new version.}.
Here are his words on the interpretation of states with negative energies:
\begin{quote}
{\sf\small ``Dirac has tried with a very keen hypothesis to overcome these difficulties. He postulates that there are in every portion of space an infinite number of electrons which fill nearly completely, in the sense of Pauli's principle, all the states of negative energy; a transition from a positive to a negative state therefore occurs very seldom since only a few negative states are unoccupied.''

}
\end{quote} 
Apparently, Fermi deems that the  success of Dirac's theory in describing the observed facts
  is a sufficient justification for its adoption.
  For this reason, he has nothing against exploring what can be learned by hypothesising that the
   Dirac sea does really exist  - in short: it works, let's use it. 
We will return to Fermi's position several times; in the meantime, let us note that we find the same attitude in his young assistants\footnote{Note that they are both twenty-five years old at the time and publish works alone on topics they learned from their renowned teacher. This tests to Fermi's attitude of respect for science and people.}
Wick and Racah, who espoused and explored Dirac's sea scheme in two works
\cite{wick} and \cite{raccoon}, both of 1934.

\bigskip
Majorana's first intervention in the discussion on \cite{m1} is
very interesting\footnote{Apparently this result was obtained during a systematic exploration of what constraints on the possible wave equations are imposed by Einstein's relativity principle.}. He wonders whether there is a theory such that, when the particle is stationary, does not have states with negative energies, as it happens to Dirac's theory. Majorana tries to formulate such a wave theory, and he finds one which, despite being {\em much} more complicated than that of Dirac,  seems to indicate that this programme  can be partially (not fully) fulfilled. 
He describes his findings in a work published in December 1932. But as we have remembered, in the same
year, Anderson provides evidence for the existence of a particle identical to the electron except for its electric charge, 
which is opposite and equal in size; a result that instead fits perfectly  in Dirac's interpretation.
The mathematical interest in the equation  discovered by Majorana is widely recognized,  but it finds no application in physics (apparently, the same fate as the Weyl equation, but we will have to return to this later).

  Interestingly, even in early 1933 many talented theoretical physicists maintained reservations about Anderson's results;
as mentioned in the appendix~\ref{app:gent}, we also find traces of this phase of the discussion in the letters that Majorana writes from Leipzig, where he is Heisenberg's guest. However, in mid-1933 scientists began to realize that the observational evidence was serious, and at the end of 1933, as we have already said, Dirac was awarded the Nobel Prize. From here on the disagreement with Dirac's ideas seems to almost disappear and very few physicists (Majorana among them) continue to reflect on it seriously.

It is not surprising that many of those who were skeptical would take Dirac's ideas on antimatter  seriously in 1934, see e.g.~\cite{dau} and~\cite{au}. Heisenberg's change of attitude is noticed by Majorana, see appendix~\ref {app:gent}, but this does not put Majorana in crisis with his intimate convictions. In short, after 1934 there were few who opposed Dirac's theoretical views; another was Pauli, who however succeeded to feel solidarity with Heisenberg. Majorana proceeds in absolute solitude.

 \medskip
One might be led to believe that Majorana emerges defeated from this first confrontation with Dirac, but on closer examination,
it can be recognised that the attempt to question such an important conclusion as the existence of antimatter, 
proceeding mathematically 
and relying on principles accepted by just about everyone  (such as relativity and wave nature of matter), testifies to Ettore Majorana's great moral integrity and exceptional talent\footnote{This talent was immediately recognized by the mathematician van der Waerden and then, five years later, by a scientific work by the physicist Eugene Wigner which recognized the foresight and value of Majorana's investigation.}.

\subsection{The nature of the neutron and the theory of the atomic nucleus}
The model of the nucleus in vogue throughout the 1920s (which can be retraced in embryonic form even before Bohr) differs from the one known to everyone today and {\em postulates that the atomic nucleus is made up of protons and electrons} - yes, electrons instead than neutrons, which had not yet been discovered. This model has the advantage of directly accounting for the possibility that some particular nuclei emit high-energy electrons or, as they were called at the time, $\beta$ rays.

A modern-day physicist immediately understands that quantum mechanics would imply that the nuclear electron, if it had existed, would have to possess a very large kinetic energy, even too large compared to the differences in mass of the nuclei. But physicists of the time (following Bohr) expected that the physics of the nucleus could radically surprise them, leading to conceptual schemes different from those of just established science of the atom - i.e., they were ready to a further intellectual revolution.

This electron-proton model of the nucleus 
{\em does not enter} into crisis even after the discovery that the emitted electrons have a variable energy: the model is only made a little more complicated so as not to contradict the observations\footnote{For readers interested in the philosophical aspects, I underline that the conceptual bases underlying these models are always those of immutable atoms. For further discussion see e.g.~\cite{milo}.}.
Proceeding along these lines, in 1930 Pauli will propose the explanation, that there is also another particle in the nucleus,
almost invisible and with very small mass: the neutrino. (The origin of its funny name is recounted in the appendix~\ref{app:nt}.)
Pauli assumes that the electron is always emitted in pair with a neutrino, and shares the available energy with it. Therefore, a portion of energy it seems to be lost, since the neutrino is very difficult to detect; in practice, it is unobservable.

Ideas in this regard will change after Chadwick's famous discovery (1932):
there is a neutral particle with a mass similar to that of the proton, but without electric charge~\cite{nobelchadwick} - the neutron. In the same year, Iwanenko first, and shortly after Heisenberg, discuss the numerous advantages of imagining that the nucleus is made only of protons and neutrons, just as we think today.  The schemes of quantum mechanics had proved applicable to both atom and nucleus: a new scientific revolution was not needed.

Majorana, still in Leipzig,  
criticises and improves the model of his famous colleague and host,
describing his proposal in a work of 1933~\cite {m2}.
Heisenberg, who had received the Nobel Prize just the year before,
recognises the point of his Italian guest and makes him famous in the world.
In this way, the 27-year-old Majorana became an internationally recognised scientist and, apparently, this success did not go unnoticed even in the place where Majorana had started out from - Rome, Via Panisperna: from this moment on, Enrico Fermi's public declarations of esteem for his collaborator's science multiplied.

\subsection{Majorana opens the doors to modern physics}\label{apuro}
To talk about Majorana's most famous work \cite{m3}, therefore moving on to the last chapter of his scientific story, it is appropriate go back to Enrico Fermi, who 90 years ago - it was 1933 - wrote one of his most important scientific papers \cite{rs33}. In that work an important question is addressed: now that we are sure that the nucleus does not contain electrons, how is it possible to describe the fact that it emits  $\beta$ rays,
i.e., high energy electrons?

Previously, as early as 1930, Ambarzumian and Iwanenko had suggested that electrons were not
extracted from the nucleus, but rather created, somehow like photons\footnote{The term Einstein originally introduced for photons was
{\em Lichtquanten,} namely, quanta of light.} can be created by an atom. The point had then been elaborated in 1933 by Francis Perrin,
who assumed that the same thing happened to neutrinos. (See \cite{milo} for references.)

These ideas arise from the vision that de Broglie's proposed in 1924, which assimilates the treatment of electrons to that of photons, and which conceives
 in terms of waves both the particles that constitute matter and those that constitute radiation. This vision included all the microscopic reality then known.
But until then, no one had been able to provide a sufficiently elaborate theory that would allow calculations and predictions to be made. Fermi's work is successful precisely because it consists of such a theory. However, to obtain this result, 
Fermi can do nothing better than rely on Dirac sea theory, as emphasised in \cite{milo}.

%
%
%In precedenza, e gi\`a nel 1930,  Ambarzumian ed Iwanenko  avevano suggerito che gli elettroni non fossero 
%estratti dal nucleo, ma piuttosto creati, un po' come i  fotoni\footnote{Il termine originariamente introdotto da Einstein per i fotoni era 
%``quanti di luce''.}   possono essere  creati da parte di un atomo; il 
%punto era stato poi elaborato nel 1933 da Francis Perrin, che aveva supposto che  ai
%neutrini capitasse lo stesso\footnote{Queste idee  nascono nell'alveo della visione di de Broglie, 
%che assimila il trattamento degli elettroni con quello  dei fotoni,   concependo  in termini di onde sia le particelle di 
%materia che quelle di radiazione.  Una tale visione include tutta la realt\`a microscopica conosciuta all'epoca.}.
%Ma  fino a quel momento, nessuno era stato in grado di fornire una teoria sufficientemente elaborata 
%che consentisse di effettuare calcoli e previsioni.  Il lavoro di Fermi ha successo, proprio perch\'e 
%consiste in  una teoria del genere. Ma per ottenere questo risultato,  Fermi non riesce a far niente di meglio che  appoggiarsi alla teoria del mare di Dirac \cite{milo}.

In the same period, a sort of ``first disappearance'' of Majorana occurred, which lasted until 1937.
Although living in Rome, he isolated himself from Fermi's group and from the world, not publishing anything for four years.

% All'incirca nello stesso periodo, avviene una specie di ``prima scomparsa'' di Majorana, che durer\`a fino al 1937.
%Pur vivendo a Roma, egli si isola dal gruppo di Fermi e dal mondo, non pubblicando pi\`u nulla per quattro anni.  

Then, in 1937, as soon as the first useful competition for a position in  theoretical physics was announced, Majorana sent  to the journal {\em Il Nuovo Cimento} a revolutionary  contribution to science - his last one.
In this  paper, he shows how we can completely avoid talking about negative energy states; in other words, we can renounce to  talk about 
{\em anti-electrons} (the term introduced by Dirac to describe the prediction of his hole theory) without having to give up talking about {\em positrons} (the term introduced by Anderson to indicate the positively charged particle, seen in experiments). The new mathematical scheme predicts only positive energy states. It describes at the same time electron waves produced by a given interaction, and positron waves absorbed by the same interaction. Appendix~\ref{app:am} elaborates the point under discussion; here it will be enough for us to note that the description proposed by Majorana is now adopted by any particle physicist. 

It is very interesting that the point of view that Majorana introduced, in addition to freeing physics from the need to hypothesize the ``Dirac sea'', allows them to conceive a special possibility for those particular matter particles that have no electric charge: if there is nothing to distinguish them, the emitted particle and the absorbed one {\em could coincide}. This hypothesis is under observational scrutiny in many laboratories around the world, including that of Gran Sasso; we will talk more about it immediately below.

Fermi's appreciation emerges from a report from the competition commission, which he chaired, for three positions in theoretical physics advertised in Palermo.
The text of the minutes~\cite{eras}
contains this statement on the achievement of Majorana
\begin{quote}
{\sf \small 
``In a recent work he has finally devised a brilliant method that allows for symmetrical treatment the positive and negative electron, eliminating at last the need to resort to the extremely artificial and unsatisfactory hypothesis of an infinitely large electric charge diffused in all space, a question that had been addressed in vain by many other scholars''

}
\end{quote}
(my translation).
While what is stated is strictly speaking correct, an allusion or implication is perhaps a little less so; it almost seems that the general thought of scientists was to get rid of the Dirac sea. Instead, as we saw above, there were also those who took advantage of this hypothesis;
compare in particular with the section~\ref{f1}, and with what was said just above.

I would like to stress that this report is not a scientific work, but a formal document aimed at achieving a very precise effect.
With this document the commission addressed Education Minister Bottai, declaring that it did not know how to proceed and asking if it was possible to assign Majorana a supernumerary post, activating the special procedure that had only been used for Marconi, after the Nobel Prize. 
Other behind-the-scenes stories, recounted by Amaldi~\cite{ama} and Sciascia~\cite{leo}, remind us how many invisible actors intervened to achieve this result - which achieves the goal of obtaining an extra place, but in a way that is difficult to consider decorous. For a look at Majorana's feelings, see appendix~\ref{app:gent}. (Maybe I should not say this but this story fills me with sadness).

For further discussion of Majorana's  scientific results, see appendices~\ref{etera} and \ref{cappon}.

\section{Actuality of Majorana's vision}\label{e-s4}
The formalism proposed  by Majorana to describe the particles of matter 
is  the basis for countless applications. 
As already mentioned, all high-energy physicists today know how to use it and use it frequently.

Since Majorana's time the understanding of interactions has improved considerably and this concerns:\\
1)~the electromagnetic forces that explain, among other things, the structure  of atoms;\\ 
2)~the strong forces, which make it possible to describe the stability of atomic nuclei; \\
3)~the weak forces, which concern the emission of $\beta$ rays and neutrinos. \\
An important specific aspect, recognised by the scientific community 
in the mid-1950s, relates  the hypothesis that neutrinos have exactly zero mass to the 
very characteristic structure of weak interactions.
This implies that the scientific progresses of fifties revamped  
the interest in the approach of Weyl, that were mentioned in section \ref{f1}. 
In those years,  Majorana's neutrino hypothesis seemed to lose interest\footnote{\label{papon} It soon became clear that if neutrinos really had 
zero mass, their spin should always be aligned with the velocity, whereas the opposite would be true for antineutrinos. 
There was a way to univocally distinguish neutrinos from antineutrinos; which apparently 
is exactly the opposite of the hypothesis 
proposed by Majorana. As we shall see, however, a distinction must be made between the case of fast-moving neutrinos and that of neutrinos at rest.}; however,  it became clear  over time  that 
the masses of neutrinos, while playing a very special role,   are not obliged to be zero, as had appeared for some time.
 Neutrinos with non-zero Majorana mass can coexist with the peculiar structure of weak interactions.
 The implication works the other way round: if the mass of neutrino is zero, their   interactions has a peculiar structure; but if the interaction has this peculiar structure, the mass of neutrino can be zero or not.   
 (To deepen this discussion, see~\cite{pal}, \cite{stuio} and 
 \cite{nata}.)

In 10-15 years, the results and ideas just mentioned become the pillars
of the so-called ``standard model'',
which is essentially a theoretical summary of all known things
about particles of matter and the forces to which they are subject - excepting gravity.

It is very interesting to consider that the only observational clue that the ``standard model'' \underline{is not} a complete theory concerns neutrinos\footnote{To date no other studies have been carried out
with particle accelerators, or with other types of apparata, has managed to provide convincing and definitive 
experimental evidence that the standard model has other fatal flaws.}.
 In the ``standard model'' neutrino masses are expected to be exactly zero; 
however, certain speculations put forward by Bruno Pontecorvo (another of the boys from via Panisperna), further refined by Sakata and his disciples in Japan, have recently found observational confirmation, leading to the conclusion that the masses of neutrinos are different from zero - as recognised by the 2015 Nobel Prize in Physics. 
We are now certain that neutrinos have a mass, much smaller than that of other matter particles, but they do. This observation contradicts the prediction of the ``standard model'', and as we have known since the time of the Greeks, the first rule of a theory is {\em sozein ta phainomena} - do not contradict observational facts.

%In 10-15 anni, i risultati e le idee appena accennate diventano i pilastri 
%del cosiddetto ``modello standard'', 
%che \`e in sostanza un riassunto  teorico di tutte le cose note
%sulle particelle di materia e le  forze a cui sono soggette.
%
%\`E molto interessante considerare che l'unico indizio osservativo  
%che il modello standard {\em non sia} una teoria completa riguarda proprio i neutrini e le loro masse\footnote{Ad oggi nessun altro studio effettuato  
%con  acceleratori di particelle o altri tipi di apparati \`e riuscito a fornire provi sperimentali convincenti 
%che il modello standard presenti falle fatali.}. 
%In effetti, certe speculazioni avanzate da 
%Bruno Pontecorvo (un altro dei ragazzi di via Panisperna), che sono state  
%precisate e raffinate da Sakata e i suoi discepoli in Giappone, hanno recentemente trovato riscontri osservativi,
%come riconosciuto dal premio Nobel in fisica del 2015. 
%Per questo, oggi siamo sicuri che i neutrini abbiano una massa diversa da zero, anche se 
%molto  pi\`u piccola di quella delle altre particelle di materia. 

 \begin{figure}[t!]
\centerline{\includegraphics[width=0.8\textwidth]{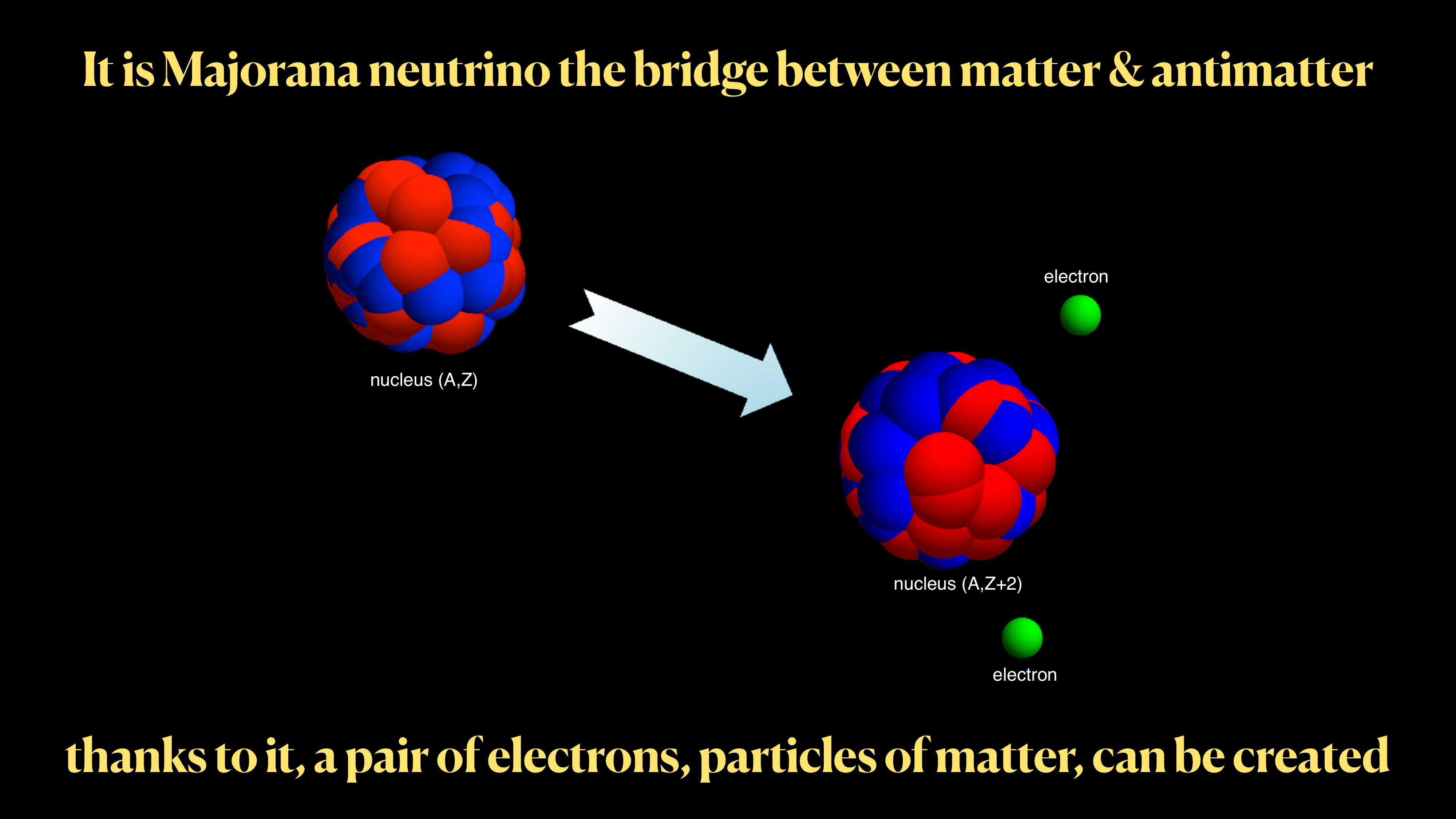}}
\caption{\small\it  Illustration of the hypothetical nuclear transformation 
 in which two of the neutrons contained within a nucleus become two protons, and 
at the same time  an electron pair is created.
This transformation 
is allowed by Majorana's neutrino theory:
the speed of the transformation 
is proportional to the square of the neutrino mass associated with the electron. 
It is the subject of passionate research in laboratories all over the world, including those at the Gran Sasso lab, which are at the forefront.
\label{fig2}}
\end{figure}

These considerations have brought Majorana's hypothesis back into vogue\footnote{The reason, in a nutshell, is that
once we are sure that neutrinos have a mass other than zero, we can consider them in their rest system, where by definition the velocity is zero.
Therefore, referring to the footnote number~\ref{papon}
we understand that - from the spin direction alone - 
there is no longer any way to distinguish neutrinos and antineutrinos.}, and 
in particular, they have motivated the search for a new potentially observable 
process~\cite{furry} that we can describe synthetically using a couple of formulas.\\
$\star$ Fermi's theory of $\beta$ ray emission convinced us that neutrons can transform spontaneously
as follows
$$
\mbox{neutron}\to \mbox{proton}+\mbox{electron}+\mbox{antineutrino}
$$
This reaction is written in modern language. It has the somewhat disturbing characteristic that particles can turn into each other. However, there are two important laws at work that make it reasonable. First of all, 
the number of heavy particles in the atomic nuclei (protons and neutrons) remains unchanged.  Moreover, we can  argue that the net number of particles has remained unchanged, because an electron has appeared but also a {\em anti}-neutrino; this can  assimilated to the algebraic situation $+1-1=0$.\\
$\star$ Now, assume with Majorana that neutrinos and antineutrinos are the same thing. Then, considering a pair of transformations like the previous one, and eliminating the neutrinos that appear in the final state, we would be led to 
consider seriously the existence of a process of this type\footnote{This transformation  is called in physicist jargon ``neutrinoless double beta decay''; see also the note number~\ref{nota18}.}. 
$$
\mbox{2 neutrons}\to \mbox{2 protons}+\mbox{2 electrons}
$$
$\star$ At this point, we should admit that something formidable can happen. In fact, we have two extra electrons at the end of the process, i.e., two previously absent particles of matter.
In more explicit terms, we would simply be witnessing a process of matter creation. 
Experiments have already established that this process, if it exists at all, is very rare; but  having assessed the credibility of the theoretical ideas that lead us to consider it,  and in view of its importance - creation of matter! -  we feel it is right to pursue it.

Numerous experiments underway in laboratories around the world are working for this goal.
   Large quantities of special chemical elements are kept under attentive observation, waiting for a transformation into a new species, whose atomic number is twice larger\footnote{For several experimental reasons, 
a promising case is the the transformation of the 
nucleus with $Z=32$ and $A=76$ into the one with $Z=34$ and $A=76$ - germanium becomes selenium.}.  
As the transformation concerns the nuclei, this has to be accompanied by the emission of two electrons, which, if observed, 
    would testify to the correctness of Majorana's hypothesis on the nature of neutrinos: 
See figure~\ref{fig2} as a summary.

%
%Numerous experiments underway in laboratories around the world are working towards this goal.
%They keep under  observation a large amount of a certain type of chemical element (e.g., the isotope of germanium with $Z=32$ and $A=76$)
%waiting for the emission of the two electrons to occur, which corresponds
%to the transformation into a chemical species with an electrical charge
%increased by two units
%  (e.g., the selenium isotope with $Z=34$ and $A=76$).
%His observation would testify to the correctness of Majorana's hypothesis on the nature of neutrinos.
%See figure~\ref{fig2} as a summary.
%
%Numerosi esperimenti in corso nei laboratori di tutto il mondo sono impegnati a questo scopo. 
%Essi tengono sotto  attenta 
%osservazione una grande quantit\`a di un certo tipo di elementi chimici (p.e., l'isotopo del germanio con $Z=32$ e $A=76$)
%in attesa che avvenga l'emissione dei due elettroni, che corrisponde 
%alla trasformazione in una specie chimica con carica  elettrica 
%accresciuta di due unit\`a  
% (p.e., l'isotopo del  selenio con $Z=34$ e $A=76$).
%La sua osservazione testimonierebbe la correttezza dell'ipotesi di Majorana sulla natura dei neutrini.
%Si veda la figura~\ref{fig2} come riassunto.

\section{Final considerations}\label{e-s5}

\begin{figure}[t!]
\centerline{\includegraphics[width=\textwidth]{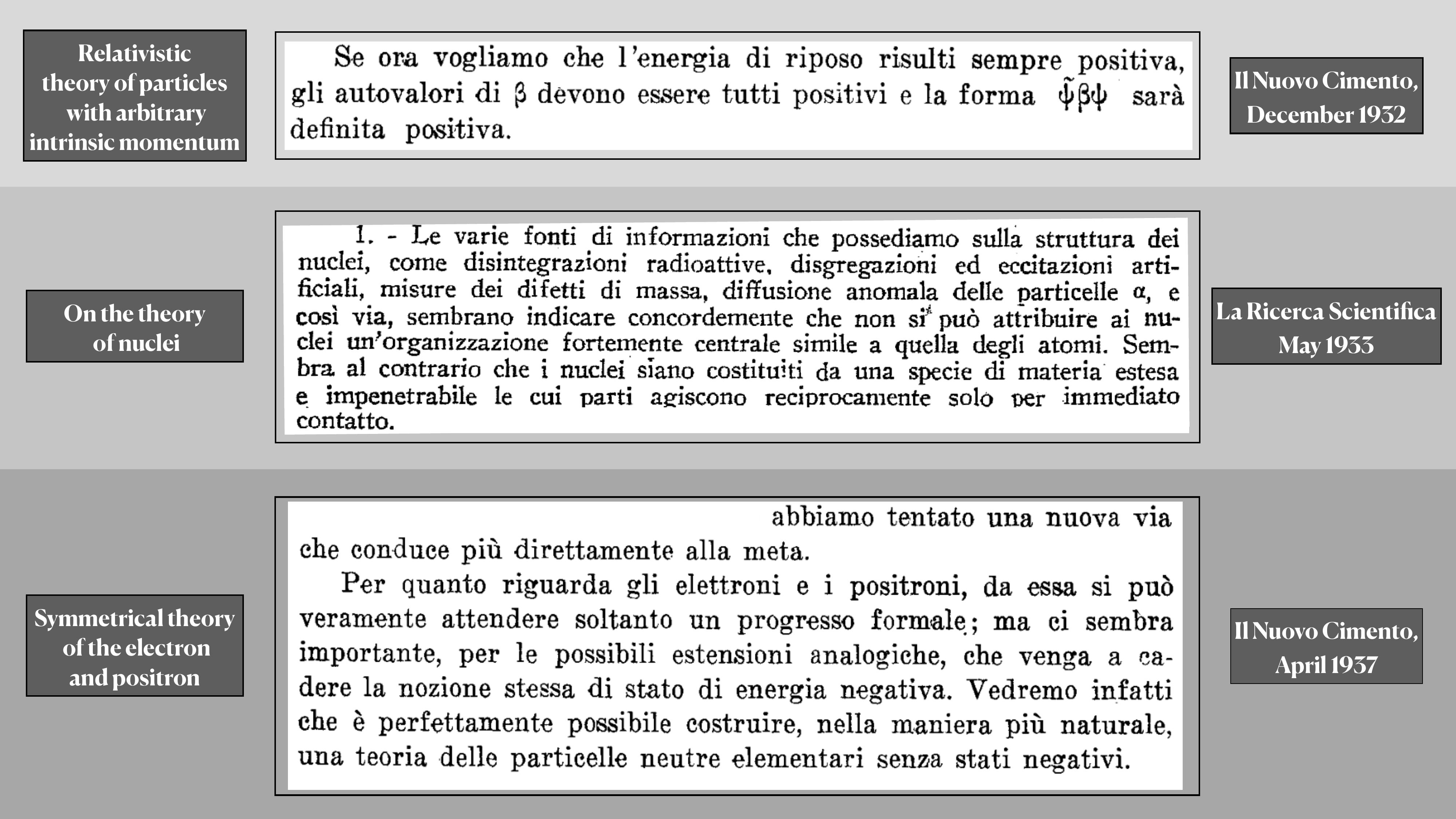}}
\caption{\small\it  Some short and significant passages from the three works by Ettore Majorana mentioned in the text.\label{fig3}}
\end{figure}

The impression is that over the years, also thanks to the work of some talented 
Italian historians, we are slowly doing 
justice to Ettore Majorana. 
His results form the basis of modern nuclei and particle science.
His neutrino hypothesis points to the future of physics: it is studied at Gran Sasso and elsewhere in the world. An attentive examination of how to proceed 
is the main objective of the conference we are about to attend.

Majorana's rigour his way of working liken him more to a mathematician than to a modern physicist, and I think this title is a great honour. My colleague Salvatore Esposito, one of the greatest experts on Majorana  
has rightly compared him to Archimedes.

I find  Majorana's  style admirable - his scientific writing are accurate, precise, elegant. The figure~\ref{fig3}
shows us short passages from the three works above discussed, that 
describe his motivations for undertaking them. Note
the use of polite and cautious expressions 
such as ``if we want to'' or ``it seems that'' or ``we have tried'' and also ``it is possible to build''.   
Note also the precision in distinguishing
between ``positrons'' and ``anti-electrons'', perhaps little understandable today, but as we have seen, 
very appropriate to deal with a hot problem of his time: how to use the Dirac equation 
without invoking the Dirac sea hypothesis. The problem has been solved so well that we can use also 
the term  {\em anti-particle} - born within Dirac sea interpretation - with very little risk of confusion. 
%(See again appendices~\ref{app:n} and \ref{etera}.)

Perhaps I am not the only one who thinks that the continuation of Majorana's personal story is, 
before being mysterious or intriguing, quite distressing. 
I do not feel prepared to talk about it, but every now and then 
I wonder if we would know how to understand, help and save exceptional people like him today.

And as we draw this discussion to a close, two words keep coming to mind. 
One is {\em courage:} it takes a lot of courage to pursue a line of research in firm opposition to a recognised genius like Dirac and also to the very strong group in which one grew up.  And another word, which I fear needs no explanation, is {\em loneliness} - too much loneliness.

\subsection*{Acknowledgements}

I would like to thank Francesco Cappuzzello, Manuela Cavallaro and Titti Agodi for the invitation and 
Salvatore Campanella of the Grimaldi Foundation for pleasant conversations and help.
I am grateful to Salvatore Esposito for valuable discussions and clarifications,
to Luisa Cifarelli for encouragement and to Francesco Cappuzzello,
Angelo Malavasi and Giulia Ricciardi for careful reading. 
With the support of  the research grant number 2022E2J4RK {\em PANTHEON: Perspectives in Astroparticle and
Neutrino THEory with Old and New messengers} 
under the program PRIN 2022 funded by the Italian Ministero dell'Universit\`a e della Ricerca (MUR) and by the European Union - Next Generation EU.
%
%Ringrazio Francesco Cappuzzello, Manuela Cavallaro e Titti Agodi per l'invito e 
%Salvatore Campanella della Fondazione Grimaldi.
%Son grato a Salvatore Esposito per preziose discussioni e chiarimenti,
%a Luisa Cifarelli per l'incoraggiamento e a Angelo Malavasi per l'attenta lettura. 
%Col supporto della borsa number 
%2022E2J4RK 
%{\em PANTHEON: Perspectives in Astroparticle and Neutrino
%THEory with Old and New Messengers}   
%parte del PRIN 2022 del Ministero dell'Universit\`a e della Ricerca. 

\appendix
\section{Appendices}\label{e-sa}

\subsection{Passages from the correspondence with Gentile Jr}\label{app:gent}

A valuable document to follow 
the evolution of Majorana's thought are his letters, very appropriately collected and presented 
by Erasmo Recami in 2000.
We highlight some passages that refer to the theme proposed in this discussion, selected from 
those addressed to his colleague and friend 
Gentile Jr - the first two sent from Leipzig in 1933, 
the third from Monteporzio (near Rome)  the last from Rome, both in 1937:
\begin{itemize}
\item[12 Mar] {\sf ``In Leipzig they had no news of the positive electron. Here they say it is a formidable quid pro quo.''}
 Note that Anderson's work is from the previous year.
 \item[7 Jun] {\sf ``People are taking Dirac's theory of positive electrons very seriously.''} Majorana refers in particular 
 to Heisenberg's change of attitude; the same opinion had also been expressed  in a letter of 22 May 1933  to Segr\'e.
 \item[25 Aug] {\sf ``I know nothing about the work I published in Nuovo Cimento except that a Swiss
 asked me for an extract. I am afraid it will spoil his holiday.'' } It is not unreasonable to assume that ``the Swiss'' was Stuecklberg, 
 who on 21 February 1938 \cite{stu38}
 will be among the first to refer to the work that  Majorana presented 
 for the 1937 professorship competition, and that was then sent to various colleagues.
 \item[21 Nov]  {\sf ``I have seen Racah's work, but only in drafts. 
 In the second part there is something real: that is, the actual application 
 to the theory $\beta$ and the criticism he addresses to me''}.
The work to which Majorana refers \cite{racah} had appeared on 15 July, three months after his own.
\end{itemize}
(my translations).  As for the last letters, 
recall that a competition had just taken place for three posts as professor of theoretical physics,
 which would have seen as winners Wick, Racah and Gentile Jr in rank order.
  Majorana had been excluded from the trio, to receive an 
  appointment ``for high and deserved fame'' (see section~\ref{apuro}).
In a letter written immediately afterwards we read
\begin{quote}
{\sf \small``I am surprised that (...) you doubt my good stomach, metaphorically speaking (...)
 if at the next conclave they make me pope on exceptional merits, I'll certainly accept it'' } 
 \end{quote}
 (my translation). 
A feeling of unease towards the unorthodox appointment procedure that had affected him, barely disguised by the veil of irony, seems to leek out.
The same letter also reveals some annoyance towards Racah\footnote{\label{nota17}In 1937, Racah 
acknowledges Majorana's new neutrino concept but does not mention the 
Dirac sea-free quantisation procedure \cite{racah}.
Majorana declares to Gentile Jr his conviction that Pauli 
was behind the choices   of their common 
colleague. Indeed, Pauli,  who  had hosted Racah between 1931 and 1932,
is thanked in written in \cite{racah}.
See also appendix~\ref{cappon}.}.

  \subsection{Names for negative and positive electrons}\label{app:n}
  Since the days of Greek mathematics, the ability to adopt new names to describe new situations 
has been a characteristic of mature science. In the case of elementary particles, one should bear in mind that 
the first evidence of the electrical nature of matter came from chemistry. The discovery 
of the electron as a particle of matter
was certified by the Nobel Prize in Physics to JJ Thomson~\cite{nobelthomson} (1906),  and that of the positron by the Nobel 
 to Anderson \cite{nobelanderson} (1936). 

\begin{table}[t]
\centerline{
\begin{tabular}{cccc}
Name & Proponent & Charge  & Remarks \\ \hline
Electron & Stoney &  $- $ &  o.m.: ion in a solution   \\
(Atom of electricity) & von Helmholtz &  $\pm$  & depends upon model \\ 
(Corpuscle) & JJ Thomson & $-$  &  matter particle \\
$\beta$  Ray & Rutherford & $- $ & high energy emission \\
(Negative energy state) & Dirac & $-$ &  interpretation of Dirac eq. \\
Anti-electron & Dirac & $+$ &  o.m.: hole in Dirac sea  \\
Positron & Anderson &    $+$ &  observed particle \\
(Negatron) & Anderson & $-$ &  proposed name  \\
%anti-particella, anti-materia &  - &  varie  & termini generici \\ 
\end{tabular}}
\caption{\em\small Names to refer to the electron or its anti-particle.
We indicate in brackets the names not in use today and highlight 
some original meanings {\rm (o.m.)}~abandoned in favour of the current one.\label{tab1}}
\end{table}

It is important to reiterate that the term {\em anti-electron} was originally a theoretical construct, 
moreover born in an interpretative context now abandoned. To 
 be sure, it is sufficient to read what Dirac said in the lecture he gave in Stockholm on the occasion of the awarding of the Nobel Prize~\cite{nobeldirac}
\begin{quote}
{\small\sf 
``We now make the assumptions that in the world as we know it, nearly all the states of negative energy for the electrons are occupied, with just one electron in each state, and that a uniform filling of all the negative-energy states is completely unobservable to us. {\em Further, any unoccupied negative-energy state, being a departure
from uniformity, is observable and is just a positron.}''

}
\end{quote}
Table~\ref{tab1} summarises various considerations presented in the text. We note {\em en passant} that,
the term $\beta$ ray has been retained in the current nomenclature to speak of electrons of relatively high energy and 
the term $\beta^+$ ray to refer to speak of positrons instead\footnote{\label{nota18} The use of the jargonic term {\em beta} certainly does not confuse
specialists but I am not sure it helps novices. Moreover, as we have mentioned, the locution ``beta ray'' was consolidated at the time when it was believed that such electrons were housed in atomic nuclei. Enrico Fermi himself distances himself somewhat from this usage, and 
in the title of his famous work  \cite{rs33}, quotes it as  {\sf raggi \guillemotleft beta\guillemotright}, i.e., in inverted commas.}.
In addition, the use of the ``anti-'' prefix has been retained, to avoid duplicating 
the number of names for all known particles, and in this way the original theory of Dirac is in a sense ``overwritten''\footnote{This may make life easier for professional scientists, but it makes history incomprehensible, which as a rule inhibits the possibility of  changes or innovations.}.

  \subsection{Origin of the word {\em neutrino}}\label{app:nt}
 The funny name of this particle 
originated in the school of physics in Rome. 
There are at least two versions, not inconsistent with each other but  
slightly different, which we give below. 
 The first, well known, is in a scientific work by Edoardo Amaldi  
devoted to neutrons \cite{ea}:
 \begin{quote}
{\sf\small ``The name \guillemotleft neutrino\guillemotright\ (a funny and grammatically incorrect contraction of \guillemotleft little neutron\guillemotright\  in Italian: neutronino) entered the international
terminology through Fermi, who started to use it sometime between the conference in Paris in July 1932 and the Solvay Conference in October 1933 where Pauli used it. The word came out in a humorous conversation at the Istituto di Via Panisperna. Fermi, Amaldi and a few others were present and Fermi was explaining Pauli's hypothesis about this \guillemotleft light neutron\guillemotright. For distinguishing this particle from the Chadwick neutron, Amaldi jokingly used this funny name, -- says Occhialini, who recalls of having shortly later told around this little story in Cambridge.''

}
\end{quote}
The other version is told 
by his wife, Ginestra Amaldi, also a physics graduate and 
above all, a talented populariser, in a book written several decades earlier  \cite{ga}:
 \begin{quote}
{\sf\small  ``When one day the Italian physicist Fermi reported on this discovery in a lecture at the University of Rome, a student asked him if Chadwick's neutron was the same neutron proposed by Pauli for the phenomenon of the emission of electrons by radioactive substances: \guillemotleft No\guillemotright,  Fermi replied, \guillemotleft Pauli's neutron is much smaller: It is a neutrino\guillemotright . This joking name has stuck.''

} 
\end{quote}
(my translation). 
A third member of the Amaldi family, their son Ugo,
in his textbook 
for high school students  
prepared with the advice of the father \cite{ugo} 
recounts the following version, 
 \begin{quote}
{\sf\small  ``It was Enrico Fermi who, following a playful pun by his student Edoardo Amaldi, first used the word 
\guillemotleft neutrino\guillemotright\ at an international congress.''
}
\end{quote}
(my translation) which does not seem to add much. 
This being the case, it does not seem possible to reconstruct the story with absolute certainty, and it is not even 
of paramount importance to do so, but perhaps Edoardo Amaldi proposed the `neutronino' version, which Fermi later contracted 
into `neutrino', or the other way round;
the only thing that is absolutely clear is that it was Fermi who used this name on an official occasion.
For the previous part of the history see \cite{milo}.

\subsection{Conceptions of anti-matter}\label{app:am}

Antimatter is also used in medicine, and so many have talked about it, that no one doubts that something like this exists in one way or another. This reasonable feeling of confidence, however, is not the same as understanding the reasoning with which physicists managed to predict its existence, which was then clarified and modified; in other words, this is not the same as answering the question
   \begin{quote}\em
How do we understand antimatter?
\end{quote}
   The problem is particularly acute in educational courses, when someone wants to explain to someone else (and for the first time) what we believe antimatter is.

%Anti-matter  is used even in medicine,   and so many have spoken about it, that  
% no one doubts that in one way or another something like this exists. This, however, is not the same as 
% understanding the argument by which physicists have managed to predict its existence and then 
%clarify their ideas as far as possible. Put in other words,  
% \begin{quote}\em 
%how do we understand anti-matter? 
%\end{quote}
% The problem is particularly  acute  in educational tracks, when one intends to explain 
% for the first time   to someone what (we think) anti-matter is.

 \subsubsection{Various approaches}
 Let us look at some of the main approaches to this question.
{\small\begin{itemize}
\item 
The Dirac approach, the main one discussed in the text, 
was based on a number of assumptions he made in order to use his wave-equation - in particular, the assumption 
that there are indeed infinite states of electrons with negative energy, but that they are harmless, being 
all occupied according to Pauli's principle\footnote{Otherwise, atomic electrons could lower their energy to infinity, and stable atoms could not exist,
as discussed in section~\ref{fishy}.}.
This approach does no longer seem as valid today as it did then,  but  it retains some didactic value and it is important to 
appreciate historical progresses.
\item 
Usually, in university courses, anti-particles are presented in one of the most difficult courses, where ideas and  mathematical elaborations are heavily mixed. Unfortunately these arguments are not easy to expound, although they have the merit of being impeccable. It is hardly ever mentioned that,  
 apart from  formal embellishments of minor significance, this is the method described in Majorana's last paper - see section~\ref{apuro}.
 \item 
A substantially identical approach to the previous one was presented by Pauli in 1941 \cite{pauli1941}, pointing out that the mysterious concept of 
negative energy can alternatively be replaced by that of 
negative frequency - the key word 
he introduced to argue in this sense being 
{\em eigenvibration}. 
\item 
An interesting and rather ambitious work by Stuekelberg in the same year \cite{stu41} 
sets out to introduce anti-particles in the context of both non-quantised and quantised theory, based on 
the simple observation that they were meanwhile observed beyond reasonable doubt.  One arrives at the curious notion  
that these are particles that propagate backwards in time; a conception that harmonises perfectly with Majorana's.
\item 
 Feynman describes his method in the most accessible possible way in a 
 lecture in honour of Dirac entitled ``The Reason for Anti-Particles'' (1986) \cite{fey}. 
This approach, just like the previous one, attempts to render the point of anti-particle ``geometrically'', i.e., in a visualisable manner.  
Waves are associated with their direction of propagation; again, they are postulated to propagate  also backwards in time (I discuss this here~\cite{feyio}).
Such ideas, schemes and concepts have proved to be particularly effective and useful for practical purposes.
 \item Finally, there is a minimal description of anti-particles  based on Dirac's wave equation and  
on the principles of quantum mechanics, which would certainly have been obvious to Majorana, and that  
I hope it may be as useful to an interested reader as it was to me in class.
I allude to this in the text and discuss it some more below. 
(Those people interested in details can find them here \cite{stuio};
to fully appreciate this approach 
it  is necessary at least to know Euler's formula and Maxwell's theory).
\end{itemize}}
In short, although the first theory that led us to the discovery of anti-matter was Dirac's, as is widely acknowledged, 
most physicists make use of Majorana's approach, which is in fact 
called the {\em canonical quantisation} of fermions and which is ``algebraic'' in nature - or alternatively but equivalently that  
of Stueckelberg-Feynman, which is more visual.  
%Figure~\ref{fig88} tries to convey various  conceptual features we use to descrive anti-matter. 

\begin{table}[t]
\begin{center}
\begin{tabular}{| l|c| |  c | r | }
\hline
\sc Maxwell  & 
Electromagnetic theory with  &  
Hole theory  & 
 \sc Dirac  \\[-0.2ex]
  1873 & 
mechanical interpretation &  
based on Dirac sea  & 
  1929 \\ \hline
\sc Hertz & 
Production of &  
Positron& 
 \sc Anderson   \\[-0.2ex] 
1880 & 
propagating waves &  
discovery  & 
 1932  \\
\hline
\sc Einstein   & 
Setting aside the  &  
Fermionic quantisation & 
\sc Majorana   \\[-0.2ex]
 1905
   & 
aether hypothesis  &  
 without Dirac sea & 
   1937 \\ \hline
\end{tabular}
\end{center}
\caption{\em\small Illustration of the parallelism between the fates of the two concepts 1)~of aether, originally 
introduced to reason about light waves, 
and 2)~of Dirac sea, introduced to reason about electron waves in 
the context of relativistic theory without contradicting the existence of atoms.}\label{sds}
\end{table}

 \subsection{The analogy between the Dirac sea and the luminiferous aether}\label{etera}
We would like to observe that there are many interesting correspondences between the step taken by Majorana towards the Dirac sea and the one taken by Einstein concerning the luminiferous aether\footnote{This hypothesis was originally introduced by Huygens in order to interpret light in terms of waves, but it is also of some importance for the original formulation of Maxwell's theory, which was based on mechanical models.}. In both cases, the most recent hypothesis completely supersedes the previous one; but note also that,  in both cases, the rejected hypothesis was nonetheless sufficient to make important advances in our knowledge of the physical universe  - anticipating the discovery of radio waves in one case and that of the positron in the other. 
 See the table~\ref{sds} as a summary.

\medskip
On the other hand, there are undeniable differences between the two situations. In the case of the aether, 
the idea survived for a long time and its abandonment took a lot of efforts, whereas in the case of the Dirac sea 
progress was rapid and occurred within a decade. Moreover, Einstein's step is at the time regarded with distrust, 
while Majorana's is very quickly accepted. 
Finally, and very 
curiously, Einstein's contribution is remembered and Majorana's is usually not. I am not convinced that the main reason is 
 that the work of Majorana originally appeared in Italian, which was not a main language of scientists.  
Perhaps it is more important to take another consideration into account: scientists, like most people, do not like to remember missteps;
in other words, one wished to forget the story of Dirac's sea hypothesis as soon as possible.
From this second point of view, Majorana did his job {\em too well}.

 \subsection{Insights into Majorana's procedure}\label{cappon}

% In 1941, Pauli reproduces Majorana's quantisation procedure
%{\em without} giving him credit for it~\cite{pauli1941}.   
%But if he had estimated that result so obvious that its author did not need a mention, 
%why did he feel obliged to describe it in writing?  
%It is interesting that Maurice Jacob (a pupil of Wick and F.~Perrin) records 
%Pauli's position but underlines  that the procedure is Majorana's~\cite{mj}.
%Recall also note~\ref{nota17} concerning Racah.} 

Up to this point, we have done everything possible to avoid mathematical niceties. In this last part, without going all the way into formalism, we try to add a few more interesting details.

\begin{figure}[t!]
\centerline{\includegraphics[width=0.7\textwidth]{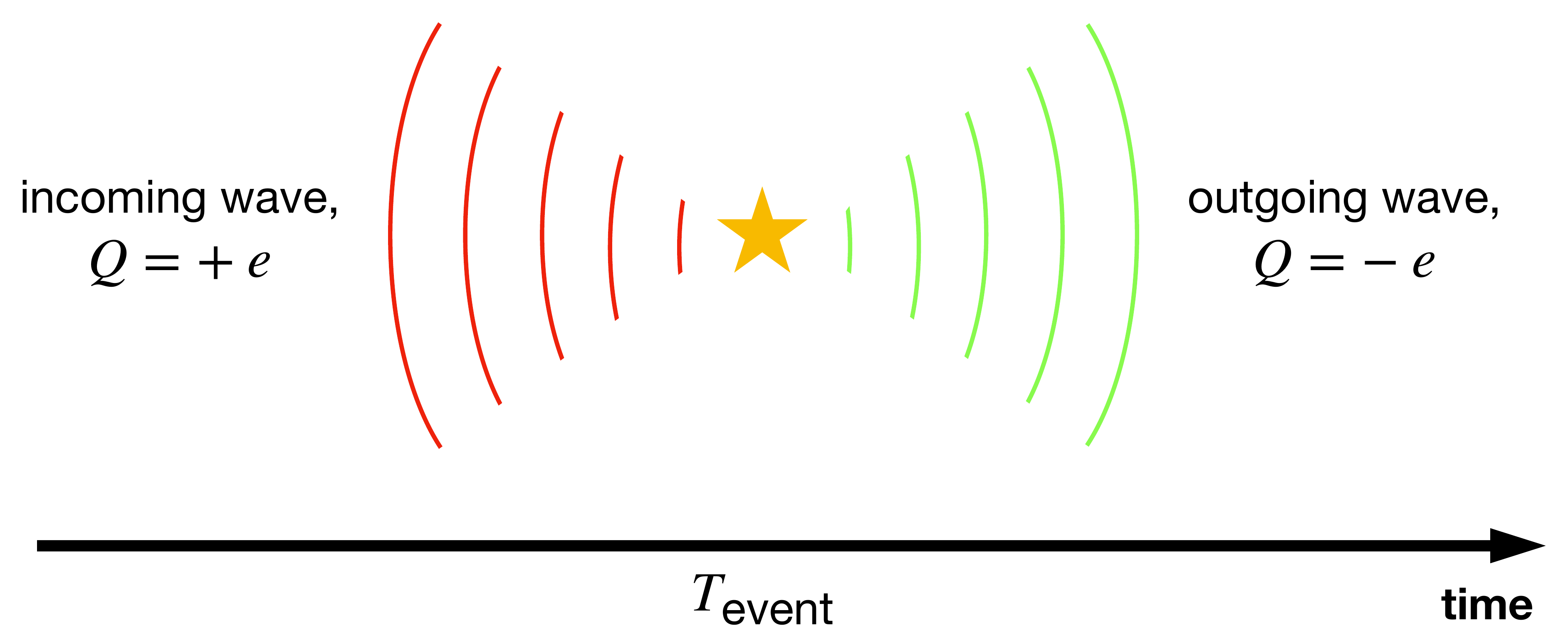}}
\caption{\small\it   In Majorana's description of  fermions and 
in any given space-time point,  
there is the  possibility of emitting   
an electron wave (depicted in green), or of absorbing a positron wave  (depicted in red). 
Imagine that  in this space-time point there is a nucleus (the star); both situations will   increase its electric charge by exactly one unit ($+e$).
There is no need to think that the nucleus gains energy by emitting negative energy electron: it is 
enough to think that the gain is due to positron absorption.
\label{fig88}}
\end{figure}

 \paragraph{The idea of Majorana expressed in the language of undulatory mechanics}
 In Majorana's theory there is no need to assume the existence of negative energy particles, as 
seemed inevitable once Dirac's interpretation had been accepted. Let us therefore better understand how this new formalism works:
\begin{quote}
Consider a system - say, a nucleus - that emits an electron with ``negative energy''  at a given time, and in this way 
{\em it
   increases its energy and electric charge.} This is the same effect caused by 
an absorption of a positron with positive energy. 
Thus, we can full avoid the problematic concept of ``negative energies'' with a symmetric treatment of electrons and positrons. We do not  
need to talk of negative energies. 
\end{quote}
This is Majorana's viewpoint. 
Note: 1) we need to introduce  a new particle in the theory to 
  entertain this interpretation, and 2) no particle will not appear or disappear alone: something else will also happen to some other particle (that above we
  have called ``the system'').
Figure~\ref{fig88} attempts to depict the specific mathematical construct, called the {\em quantised fermionic field},
that was introduced
 by Majorana to describe electrons and other particles of matter within previously described conceptual lines.

Note the analogy with what happens in atomic systems: the possibility that an atom absorbs  a quantum of light (=a photon) and in this way goes in a higher energy state, is  accompanied by the possibility of the excited atom emits a photon. 
The differences lie in the fact that the photon is a particle of light  while an electron is a particle of matter. Moreover, the photon  has no electric charge and coincides with its own antiparticle.
 As we have argued in the text, this last characteristic could assimilate the photon and the neutrino, as hypothesised by Majorana - an assumption  that harmonises well with the current understanding of particle physics and which is under intense experimental scrutiny (see section~\ref{e-s4}).

%Note the analogy with what happens in atomic systems: the absorption of a quantum of light (or photon) from an atom 
%is always accompanied by the possibility that the same atom may emit a photon. 
%The differences 
%consist in the fact that the photon is a particle of light and not of matter, 
%as electrons or neutrinos; 
%moreover,  it does not have an electric charge and 
%it does coincide with its own anti-particle.
% As we have argued in the text, this last characteristic could  
%assimilate the photon and the neutrino, assuming that the hypothesis put forward by Majorana is correct:
%this fits well current understanding of particles.
 
 To conclude the argument, let us see how this works in practice,  
considering as an example the process that today is called $\beta^+$ emission.
 In the formalism that adopts  Dirac sea - the one originally adopted to predict this process \cite{wick} - 
 the capture of a negative energy electron decreases the charge of a nucleus by one unit 
 and its energy;  in the formalism of Majorana, the {\em same} result is obtained by the emission of a positive energy positron - the advantage 
 being that in this way negative energy particles are not anymore needed. Today, when we write
 $(A,Z)\to (A,Z-1)+e^+ +\nu_e$ we have in mind the second interpretation.

 \paragraph{The choice of gamma matrices}
 As pointed out by Majorana, a specific choice of gamma matrices facilitates the presentation of his alternative treatment of 
Dirac's equation that avoids negative energies - i.e., {\em canonical quantisation.}

Let us add, in fairness, that  this choice of gamma matrices was probably already known to Dirac 
from his very first works (1920s) and to Pauli, who wrote a paper 
on the characteristics of gamma matrices in 1936~\cite{pauli36}. One could believe that Pauli, 
after his work with Weisskopf in 1934 on the quantisation of particles without spin~\cite{pw},
was close to proceed with the 
analysis of the fermions.

However, the first person in the world to describe how to avoid using the Dirac sea, when dealing with 
electrons and other fermions with half-integer spin, was Ettore Majorana in 1937 \cite{m3}.

In 1941, Pauli reproduced Majorana's quantisation procedure {\em without} giving him credit for it~\cite{pauli1941}.   
But if he considered that result so obvious that its author did not need to be mentioned, why did he feel compelled to describe it in writing?  
This choice is hard to understand and does not seem easily defensible with the claim to deal with the general form of Dirac matrices, without using a specific representation of them;  recall also   note~\ref{nota17}.
Today, it is rare to cite references on who introduced canonical quantisation of fermions. However, it is  
interesting that Maurice Jacob (a pupil of Wick and F.~Perrin) 
reports Pauli's position but emphasises that the procedure we adopt today is the one described by Majorana
%reports Pauli's position but emphasises that the procedure is that of Majorana 
\cite{mj}.

%\subsection{Una presentazione del punto di vista di Majorana}
%\subsection{}

%\newpage
%\printendnotes

%
%\newpage
%\tableofcontents

\end{document}